\RequirePackage[2020-02-02]{latexrelease}
\documentclass[9pt,twocolumn,twoside]{osajnl}
\usepackage{babel}
\journal{ol}
\setboolean{shortarticle}{true} 

\title{Square wave generation in vertical external-cavity Kerr-Gires-Tournois interferometers}
\author[1]{Elias R. Koch}
\author[1,2]{Thomas G. Seidel}
\author[1,2,3]{Svetlana V. Gurevich}
\author[2*]{Julien Javaloyes}

\affil[1]{Institute for Theoretical Physics, University of M\"unster, Wilhelm-Klemm-Str. 9, 48149 M\"unster, Germany}
\affil[2]{Departament de F\'{\i}sica, Universitat de les Illes Balears \& IAC-3, Cra.\,\,de
	Valldemossa, km 7.5, E-07122 Palma de Mallorca, Spain}
\affil[3]{Center for Nonlinear Science (CeNoS), University of M\"unster, Corrensstraße 2, 48149 M\"unster, Germany}

\affil[*]{Corresponding author: Julien.Javaloyes@uib.es}
\graphicspath{{./}{./figs}} 

\begin{abstract}
We study theoretically the mechanisms of square-wave (SW) formation in vertical external-cavity Kerr-Gires-Tournois interferometers in presence of anti-resonant injection. We provide simple analytical approximations for their plateau intensities and for the conditions of their emergence. We demonstrate that SWs may appear via a homoclinic snaking scenario, leading to the formation of complex-shaped multistable SW solutions. The resulting SWs can host localized structures and robust bound-states.
\end{abstract}

\setboolean{displaycopyright}{true}

\begin{document}

\maketitle

The generation of square-waves (SWs) in optical and optoelectronic systems has attracted a lot of attention over the past years not only for its fundamental interest but also motivated by applications such as optical clocks in signal processing, communication systems~\cite{KAA-PTL-10,SWX-OL-13} and optical sensing~\cite{USN-OE-11}. The formation of SWs due to the presence of time-delayed feedback loops was reported in optoelectronic oscillators~\cite{PJC-PRE-09,WED-PRE-12,MCE-PRE-15}, injected semiconductor lasers~\cite{JAH-PRL-15}, vertical-cavity surface-emitting lasers (VCSELs)~\cite{MGJ-PRA-07,SGP-PRA-12,MJB-PRA-13}, edge-emitting diode lasers~\cite{GES-OL-06,FVD-OL-14,IKV-OL-21}, and semiconductor ring lasers~\cite{LLZ-OL-16,MvdSG-OE-12}. 
In the framework of time-delayed systems \cite{Kolmanovskii-BOOK}, SWs typically appear in the long-delay limit via a supercritical Andronov-Hopf (AH) bifurcation as modulated oscillations with a period close to twice the value of the time-delay~\cite{CMP-83,MPN-AMPA-86,ELLG-PhysD-04,N-PRE-04}. These nascent SWs then evolve into sharp transition layers connecting plateaus of duration $\sim\tau$. 
However, SWs with other periods and different plateau lengths can also be observed~\cite{WED-PRE-12,FVD-OL-14}. 
\begin{figure}[b!]
	\centering
	\includegraphics[width=1\columnwidth]{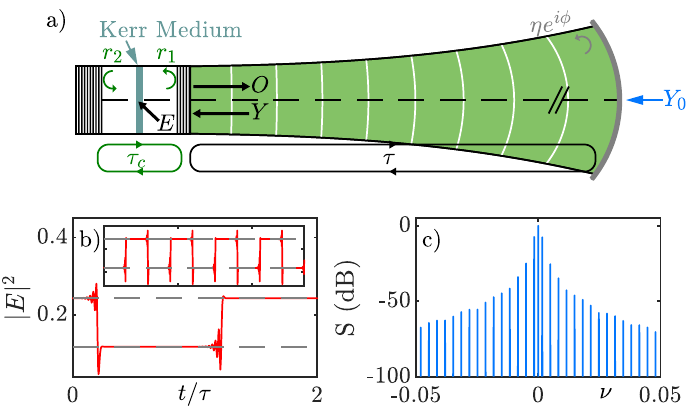}
	\caption{a) Schematic of a micro-cavity containing a Kerr medium coupled to an external cavity which is closed by a mirror with reflectivity $\eta$ and phase $\phi$, and driven by a CW beam with amplitude $Y_0$. b) Numerical simulation of Eqs.~(\ref{eq:E},\ref{eq:Y}) showing a periodic SW regime. The gray dashed lines correspond to the value of the lower and upper plateaus. The inset shows the dynamics over several round-trips. c) Power density spectrum $S$ of the solution in b) scaled to 0 dB. The parameters are $(Y_0,\delta,h,\eta,\varphi,\tau)=(0.62,0.2,2,0.9,\pi,300)$.}
	\label{fig:setup}
\end{figure} 

Recently, the existence of tunable optical frequency combs and temporal localized states (TLSs) was predicted in the dynamics of a monomode micro-cavity containing a Kerr medium coupled to a long external feedback cavity under continuous wave (CW) injection~\cite{SPV-OL-19,SJG-OL-22}. In the normal dispersion regime, dark and bright TLSs appear via the locking of domain walls connecting the high and low CW intensity levels of the bistable injected cavity. Due to the oscillatory tails induced by the cavity dispersion, these TLSs can interlock at multiple equilibrium distances leading to a rich ensemble of multistable solutions.

In this Letter, we demonstrate that the same optical system operated out of the bistability regime can leads to the formation of SWs in the multi-GHz range when the wavelength of the optical injection is set in-between two modes of the external cavity. The SWs appear as periodic solutions with a repetition rate that is approximately twice the external cavity round-trip, see Fig.~\ref{fig:setup}~(b). Using a first-principle model based on delay-algebraic equations (DAEs) we predict the SW plateau intensities as well as their bifurcation points analytically. Employing a combination of numerical simulations and path-continuation, we show, for the first time to our knowledge, that SWs may appear trough a homoclinic snaking scenario leading to the formation of complex-shaped multistable SW solutions. Finally, we demonstrate that SWs can be used as a platform to host other type of wave-forms such as robust bound states and TLSs.

A schematic setup is depicted in Fig.~\ref{fig:setup}~(a), see also~\cite{SPV-OL-19} for the details. It is composed of a monomode micro-cavity of a few micrometer with a radius up to 100\,µm and with round-trip time $\tau_c$ which contains a nonlinear Kerr medium such as silicon nitride and that is closed by two distributed Bragg mirrors with reflectivities $r _{1,2}$. The transverse degrees of freedom permit spreading the power density over a wide beam waist while still allowing for efficient fiber coupling. The micro-cavity is coupled to a long external cavity of (typically) a few centimeters with round-trip time $\tau$. This system is subjected to CW injection with amplitude $Y_0$ and frequency $\omega_0$. The detuning with respect to the micro-cavity resonance $\omega_c$ is $\delta=\omega_c-\omega_0$. The external cavity is closed by a feedback mirror with reflectivity $\eta$. The total external cavity phase $\varphi=\omega_0\tau+\phi$ consists of the accumulated phase per round-trip due to propagation and the phase of the feedback mirror $\phi$. Following the methods developed in ~\cite{MB-JQE-05,SPV-OL-19,SJG-OL-22}, the first-principle model for the system presented in Fig.~\ref{fig:setup}~(a) is 
\begin{align}
    \dot{E}&=\left[i\left(|E|^2-\delta\right)-1\right]E+hY\,,\label{eq:E}\\
    Y&=\eta e^{i\varphi}\left(E\left(t-\tau\right)-Y\left(t-\tau\right)\right)+\sqrt{1-\eta^2}Y_0\,. \label{eq:Y}
\end{align}
Here,  $E$ and $Y$ denote the slowly varying envelope of electric fields in the micro-cavity and the external cavity, respectively. Equations~(\ref{eq:E}, \ref{eq:Y}) were derived by solving exactly the field equations in the linear parts of the micro-cavity, connecting the fields at the interface with the nonlinear medium as a boundary condition. The light coupling efficiency in the cavity is given by the factor $h=h(r_1,\,r_2)$.  Here, we consider the case of a perfectly reflecting bottom mirror, i.e., $h(r_1,1) = 2$, the so-called Gires–Tournois interferometer regime~\cite{GT-CRA-64}. These cavities are theoretically lossless since all the photon are eventually reflected. They are used for inducing a controllable amount of second-order dispersion, and, using red or blue detuning, one can achieve either normal or anomalous dispersion. Around resonance, the third-order dispersion (TOD) becomes the leading term as the second-order contribution vanishes and switches sign~\cite{SCM-PRL-19}. The field cavity enhancement can be conveniently scaled out using the Stokes relations allowing $E$ and $Y$ to be of the same order of magnitude. This leads to a simple input-output relation $O=E-Y$. Note that the coupling between the fields $E$ and $Y$ is given by the DAE (\ref{eq:Y}). The latter takes into account all the multiple reflections in a possibly high finesse external cavity for which $\eta \lesssim 1$.

The feedback phase $\varphi$ corresponds to the detuning with respect to the nearest external cavity mode. In~\cite{SPV-OL-19,SJG-OL-22} it was shown that for $\varphi=0$, the Kerr nonlinearity causes a bistable CW response in a certain range of $Y_0$ and $\delta$. However, for other feedback phases, the bistability can be lost. It is the case for the anti-resonant situation in which the injection is set exactly in-between two external cavity modes. There, $\varphi=\pi$ and the intensity of the CW solution $|E(t)|^2=I_s$ can be obtained from Eqs.~(\ref{eq:E},\ref{eq:Y}) by the implicit relation
\begin{align}
    Y_0^2&=\frac{1}{k_0}\left(k_-^2+\left(I_s-\delta\right)^2\right)I_s\,, \label{eq_37}
\end{align}
while the value of $Y(t)=Y_s$ is deduced from Eq.~(\ref{eq:E}) and where we defined $k_{\pm}=\dfrac{1\pm\eta(1-h)}{1\pm\eta}$ and $k_0=h\sqrt{\dfrac{1+\eta}{1-\eta}}$. 

\begin{figure}[t!]
	\centering
	\includegraphics[width=1\columnwidth]{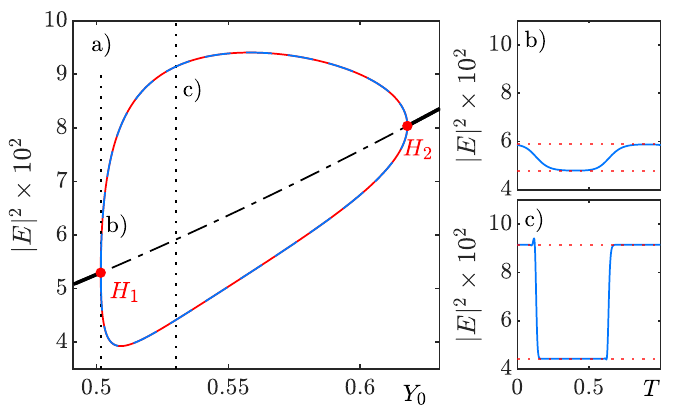}
	\caption{(a) CW branch (black) defined by Eq.~(\ref{eq_37}) as a function of $Y_0$ plotted together with the SW solutions obtained from (i) Eq.~(\ref{eq_Y0_sw}) (red) and (ii) by means of numerical path-continuation (blue). The pitchfork bifurcation points $H_{1,2}$ obtained from Eq.~(\ref{eq:BifPoints}) correspond to AH bifurcations on the CW branch giving rise to the SWs. (b) A SW profile close to the AH bifurcation $H_1$ with period $T\approx 2.013 \tau$. (c) A fully developed square wave. The parameters are $(\delta,\,h,\,\eta,\,\varphi,\,\tau)=(0.1,\,2,\,0.9,\,\pi,\,300)$.}
	\label{fig:2}
\end{figure}

One can show that $Y_0$ as given by Eq.~(\ref{eq_37}) is a monotonous function of $I_s$ thereby excluding bistability for the anti-resonant case.
However, in this regime, trains of SWs with periodicity $\gtrsim 2\tau$ can be generated as shown in the inset of Fig.~\ref{fig:setup}~(b). Note that due to the TOD induced by the micro-cavity, the resulting SWs possess oscillatory tails around both plateaus close to the transition layer. Hence, the envelope of the resulting frequency comb is asymmetrical and contains only odd frequency components, see Fig.~\ref{fig:setup}~(c).    
The mechanism at work is similar to the one observed by Ikeda et al.~\cite{IKA-PRL-82}. The latter considers the action of the cavity as a round-trip map in which CW solutions appear as fixed points. Further, SWs may appear as period-doubling (flip) bifurcations. Hence, neglecting the sharp transitions connecting the two plateaus (cf. gray dotted lines in Fig.~\ref{fig:setup}~(b)), we can approximate the SW solution as $\left(E,Y\right)\left(t\right)=(E_{1},Y_{1})$ for $t\in[0,\tau[$ and $\left(E,Y\right)\left(t\right)=(E_{2},Y_{2})$ for $t\in[\tau,2\tau[$. Heuristically, one may consider that when the system is injected during a time $\sim \tau$ with the fields whose values are given by the first plateau, it generates the response for the second one, and vice versa, which explains why the periodicity is $\sim 2\tau$.
Inserting this ansatz into Eqs.~(\ref{eq:E},\ref{eq:Y}) yields four equations for the unknowns $E_{1,2}$ and $Y_{1,2}$. Factoring out the CW solution, one can calculate the intensity of the two plateaus $I_{1,2}=|E_{1,2}|^2$ as
\begin{align}
    I_{1,2}=\left(2\delta-I_{2,1}\pm\sqrt{4\delta\,I_{2,1}-3\,I_{2,1}^2-4k_+^2}\right)/2 \label{eq_40}.
\end{align}
The Eq.~(\ref{eq_40}) indicates that for any inserted plateau intensity value, there exists two solutions for the other one. This can be explained by tracing an horizontal line in Figure ~\ref{fig:2}; each plateau intensity occurs twice, for two different values of $Y_0$. The upper and lower extremes of the bubble of these periodic solutions correspond to the case of the double root in Eq.~(\ref{eq_40}). One can construct an implicit relation between $Y_0$ and the two plateau intensities as 
 \begin{align}
    Y_0^2=\frac{I_1}{2k_0}\left|k_{-}-i\left(I_1-\delta\right)+\left[k_--i\left(I_2-\delta\right)\right]\frac{k_+-i(I_1-\delta)}{k_+-i(I_2-\delta)}\right|^2\,. \label{eq_Y0_sw}
\end{align}
Figure~\ref{fig:2}~(a) shows the resulting plateau values as a function of $Y_0$ (red line). One can see that for increasing (decreasing) $Y_0$ they emerge from the CW solution defined by Eq.~(\ref{eq_37}) (black line) in a supercritical pitchfork bifurcation. These bifurcation points $H_{1,2}$ are obtained solving Eq.~(\ref{eq_40}) for $I_1=I_2$, yielding 
\begin{align}
I_{H_{1,2}}=\left(2\delta\pm\sqrt{\delta^2-3k_+^2}\right)/3\,. \label{eq:BifPoints}
\end{align}
Remarkably, the agreement between the analytically calculated plateaus and the corresponding SW plateau values found by path-continuation of the DAE system Eqs.~(\ref{eq:E},\ref{eq:Y}) (cf. blue lines in Fig.~\ref{fig:2}~(a)) is extremely good. For the numerical path-continuation, a DAE extension~\cite{HGJ-OL-21,SJG-OL-22} of the DDE-BIFTOOL package~\cite{DDEBT} was employed. Note that the pitchfork bifurcation points $H_{1,2}$ for the plateaus correspond to AH bifurcations for the CW state giving rise to the SW solutions. The Figs.~\ref{fig:2}~(b,c) show two exemplary SW profiles for two different values of $Y_0$ (c.f. dashed vertical lines in panel (a)): a small-amplitude sinus-like periodic solution with a period $\gtrsim2\tau$ close to the bifurcation point (cf. panel (b)) transforms into a well-developed SW with sharp transition layers (see panel (c)) when $Y_0$ is increased.
\begin{figure}[t!]
	\centering
	\includegraphics[width=1\columnwidth]{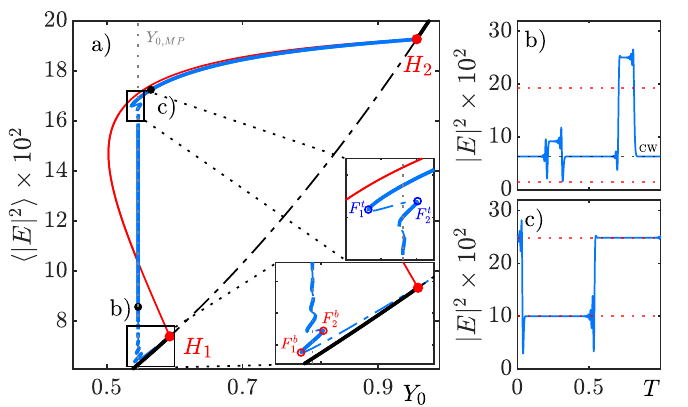}
	\caption{a) A branch of SW solutions as a function of $Y_0$ obtained by path-continuation of Eqs.~(\ref{eq:E},\ref{eq:Y}) for $\delta=0.2$ : A CW solution (solid black) becomes unstable (dash-dotted black) at a point $H_1$ in a subcritical AH bifurcation. The resulting SW solution (blue) experiences a series of saddle-node bifurcations building a collapsed snaking structure around a Maxwell line $Y_{0,MP}\approx0.54625$. Finally, the SW branch disappears in a supercritical AH bifurcation at the point $H_2$.
	The inset shows the stable (solid) and unstable (dash-dotted) SW solutions close to the $H_1$ point. The red solid line represents the average of the two plateaus, obtained analytically by Eq.~(\ref{eq_Y0_sw}). (b,c) Exemplary SW profiles marked in a). The SW solution possessing a complex shape with non-equidistant plateaus along the snaking region (b), results in a SW with a 50$\%$ duty cycle (c). See Visualization 1 for a video with the solution profiles along the whole branch. Parameters as in Fig.~\ref{fig:2}.}
	\label{fig:3}
\end{figure} 

Interestingly, this well-established scenario of the SW formation via a supecritical AH bifurcation changes dramatically when the the detuning $\delta$ is increased, see Fig.~\ref{fig:3}~(a). Here, one observes that while the AH point $H_2$ for the higher $Y_0$ value remains supercritical, the AH bifurcation at $H_1$ becomes \emph{subcritical} and the SW branch experiences a sequence of saddle-node bifurcations (see the inset of Fig.~\ref{fig:3}~(a)) resulting in a collapsed snaking structure. In particular, close to the bifurcation point $H_1$, two pairs of unstable fronts emerge from the CW solution, which is stable in the subcritical region. Here, one kink-antikink pair connects the stable CW state with the lowest of the two emerging plateaus, while another pair links the CW with the second plateau. Then these opposed fronts for each pair move away from each other. However, due to the oscillatory tails induced by TOD, the fronts can lock at several positions in the vicinity of the Maxwell line~\cite{BK_Chaos_07} $Y_0=Y_{0,MP}$, where the two fronts have the same speed and their dynamics is arrested, see Fig.~\ref{fig:3}~(b). Along the $Y_{0,MP}$ line, the front pairs gradually grow approaching each other until they finally meet at the upper snaking fold. As a result, a SW between two plateaus with a 50$\%$ duty cycle is formed, see Fig.~\ref{fig:3}~(c). For the increasing $Y_0$, the maximal intensity of the SW solution decreases and finally disappears in a supercritical AH bifurcation at $H_2$. Note that the two plateaus approximation Eq.~(\ref{eq_Y0_sw}) is valid only in the high-power region between $H_2$ bifurcation point and the upper snaking fold of the SW branch and fails in the snaking region (see the red line in Fig.~\ref{fig:3}~(a)). However, not only the position of the supercritical bifurcation point $H_{2}$, but also the subcritical bifurcation point $H_{1}$ is predicted correctly by Eq.~(\ref{eq:BifPoints}).
%
\begin{figure}[t!]
	\centering
	\includegraphics[width=1\columnwidth]{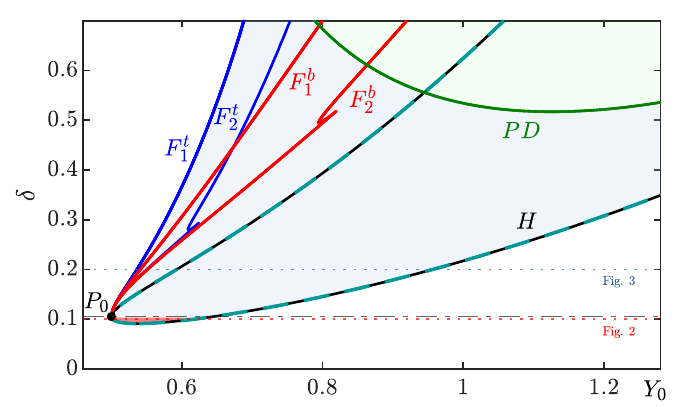}
	\caption{Two-parameter bifurcation diagram in the ($Y_0,\,\delta$) plane. Black (turquoise) lines $H$ corresponds to the threshold of the AH bifurcation obtained numerically (analytically by Eq.~\ref{eq:BifPoints}). $P_0=(Y_0^c,\,\delta_c)\approx(0.5, 0.377)$ denotes the folding point of the $H$ line, separating the super-(light red) and subcritical (light blue) regimes. Two horizontal cuts correspond to the $\delta$ values in Figs. \ref{fig:2},\ref{fig:3}. Blue and red lines stand for the thresholds of the fundamental fold pairs $F_{1,2}^b$, $F_{1,2}^t$ in the snaking region. Green line $PD$ corresponds to the onset of the period-doubling bifurcation. Other parameters as in Figs.~\ref{fig:2},~\ref{fig:3}.}
	\label{fig:4}
\end{figure} 

For a better understanding of the SW formation scenario and the transition from a super- to a subcritical AH bifurcation, it is instructive to consider the two parameter plane $(Y_0,\,\delta)$ and follow the bifurcation points using two-parameter continuations, see Fig.~\ref{fig:4}. For increasing $\delta$, the existence region of SWs increases and the $H$ curve possesses a folding point $P_0=(Y_0^c,\,\delta_c)$. Remarkably, at this point the fundamental fold pairs $F_{1,2}^b$ and $F_{1,2}^t$ emerge (cf. Fig.~\ref{fig:3}~(a)) and the snaking structure appears. Hence, the line $\delta=\delta_c$ (cf. dashed-dotted gray line in Fig.~\ref{fig:4}) crossing $P_0$ corresponds to the transition from a super- to a subcritical AH bifurcation for the SWs (cf. two horizontal cuts below and above the line corresponding to Figs.~\ref{fig:2} and \ref{fig:3}, respectively). In the subcritical regime, the distance between  $F_{1,2}^t$ and $F_{1,2}^b$ increases with $\delta$ making the snaking region wider. Note that the position of $P_0$ also depends on $\eta$. In particular, it moves into the region of smaller injections and detunings if $\eta$ is increased. Further, the SWs dynamics becomes more complex with increasing $\delta$ and, in particular, a period doubling bifurcation $PD$ sets in at higher detunings (see the green line in Fig.~\ref{fig:4}). Close above this line, for fixed values of $\delta$, SW solutions become unstable and a supercritical PD bifurcation bridge appears if $Y_0$ is increased (decreased). However, further increase in $\delta$ leads to a sequence of further PD bifurcations leading to complex chaotic dynamics. Note that chaotic dynamics for SWs was observed for e.g., delayed electro-optical systems~\cite{KCL-PRL-05,PJC-PRE-09}.
\begin{figure}[t!]
	\centering
	\includegraphics[width=1\columnwidth]{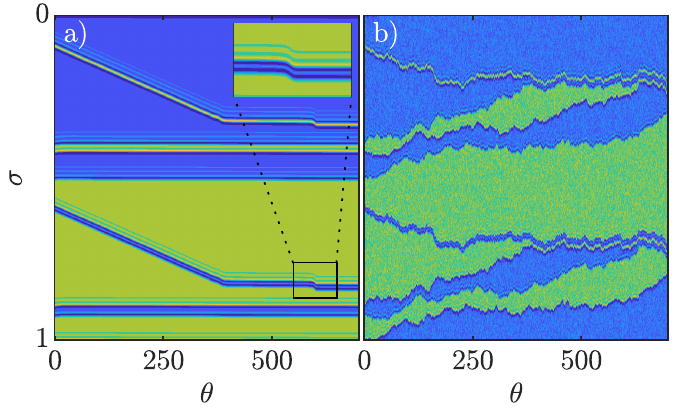}
	\caption{(a) A two-time representation of the intensity of the $E$ field obtained by numerical simulationd of Eqs.~(\ref{eq:E},\ref{eq:Y}) for $(Y_0,\delta)=(0.62,0.2)$. Two pulses of different widths propagate over the SW plateaus and build stable bound states at several preferred distances. The inset shows a zoom into the region where a transition between two bound states appears. (b) The same simulation in the presence of white Gaussian additive noise of an amplitude of 0.01. Other parameters as in Figs.~\ref{fig:2}--\ref{fig:4}.}
	\label{fig:5}
\end{figure} 

The presence of the highly oscillatory tails of SWs opens up the possibility to build more complex patterns. As an example, we integrated numerically the Eqs.~(\ref{eq:E},\ref{eq:Y}) employing a semi-implicit scheme starting with a SW solution (cf. the high power SW branch in Fig.~\ref{fig:3}~(a)) and wrote two TLSs of different sizes on their plateaus. To visualize the evolution of the resulting solution over many round-trips, a two-time representation~\cite{AGL-PRA-92} is employed. Here, the dynamics is separated into two timescales: The fast timescale $\sigma$ governs the dynamics within one period, whereas the slow scale $\theta$ describes the dynamics from one round-trip to the next one. The resulting temporal time trace is then folded with a period $\sim 2\tau$, and it is presented in Fig.~\ref{fig:5}~(a). One can see that the two TLSs start to develop oscillatory tails and propagate over the plateaus. Because of their different pulse widths, they experience distinct drifting speeds and their relative distance decreases with time. However, their slowly decaying oscillatory tails induce a series of preferred distances, where the TLSs can lock to each other forming multiple bound states (see the inset in Fig.~\ref{fig:5}~(a) around the region where the transition between two bound states with different relative positions occurs). The scenario of the bound state formation is robust against the noise, see Fig.~\ref{fig:5}~(b). Here, one can see that the presence of white Gaussian additive noise leads to a slow coarsening dynamics resulting in a drift of both the SW and the written pulses. However, the pulse interaction scenario remains unaffected and bound states are formed and broken over a slow time scale that corresponds to many round-trips.

In conclusion, we unveiled the mechanisms responsible for the formation of SWs in vertical external-cavity Kerr-Gires–Tournois interferometers. Using a delay algebraic equations model we provided an analytical approximation of the corresponding plateau intensities and bifurcation points. We have demonstrated that beyond the well-established supercritical scenario for their emergence, SWs can also exhibit a homoclinic snaking leading to the formation of complex-shaped multistable solutions. We elucidated the transition between both regimes and the parameters at which it occurs. Finally we have shown that SWs can be employed as a platform to host more complex temporal structures and we revealed that bound states of localized pulses can be formed on the top of the SW plateaus. 

$\,$

\small
\textbf{Funding}  Studienstiftung des Deutschen Volkes; Ministerio de Economía y Competitividad (PGC2018-099637-B-100 AEI/FEDER UE).
$\,$

\textbf{Disclosures} The authors declare no conflicts of interest
$\,$

\textbf{Data availability} Data underlying the results presented in this paper are not publicly 
available  at this time but may be obtained from the authors upon reasonable request.
$\,$
%
%
%

\end{document}